\documentstyle[preprint,aps,epsf]{revtex}

\newcommand{\matrxd}[4]{\left(\begin{array}{cc}  #1 & #2 \\
                        #3 & #4 \end{array}\right)}

\begin{document}
\def\btt#1{{\tt$\backslash$#1}}
\draft
\preprint{KANAZAWA 96-13,  
\ August 1996}  
\title{
Critical exponents and abelian dominance in  $SU(2)$ QCD
}
\author{
Shinji Ejiri$^{\small a}$,
Shun-ichi Kitahara$^{\small b}$,
Tsuneo Suzuki$^{\small a}$ 
\footnote{ E-mail address:suzuki@hep.s.kanazawa-u.ac.jp}
 and Koji Yasuta$^{\small a}$
}
\address{$^{\small a}$
Department of Physics, Kanazawa University, Kanazawa 920-11, Japan
}
\address{$^{\small b}$
Jumonji University, Niiza, Saitama 352, Japan
}
\maketitle

\begin{abstract}
The critical properties of the abelian Polyakov loop 
and the Polyakov loop in terms of Dirac string are studied 
in finite temperature abelian projected $SU(2)$ QCD.
We evaluate the critical point and the critical exponents from 
each Polyakov loop in the maximally abelian gauge 
using the finite-size scaling analysis.
Abelian dominance in this case is proved quantitatively.
The critical point of each abelian Polyakov loop is  
equal to that of the non-abelian Polyakov loop 
within the statistical errors. Also, the critical exponents 
are in good agreement with those from non-abelian Polyakov loops.
\end{abstract}
\newpage

\section{Introduction} %---------------------------------------

Abelian projected QCD has been studied extensively in recent years,
for elucidating the mechanism of quark confinement
\cite{suzu93,polik}.
The abelian projection of QCD\cite{thooft} 
is to perform a partial gauge-fixing  
such that the maximal abelian torus group remains unbroken.
Abelian monopoles appear as a topological quantity 
in such a partial gauge fixing,
so that QCD can be regarded as an abelian theory 
with electric charges and monopoles.
't Hooft conjectured that 
if the monopoles made Bose condensation, 
quarks could be confined due to dual Meissner effect\cite{thooft}.

There are some evidences on lattices 
that the abelian theory in the maximally abelian (MA) gauge\cite{kron}
well represents the long range properties of QCD:

\begin{enumerate}
\item
Abelian Wilson loops composed of abelian link fields alone 
can reproduce the full ($SU(2)$) value of the string tension.
Furthermore, the abelian Wilson loops written in terms of 
monopole currents also reproduce the value
\cite{yotsu,hio,shiba3,shiba4,ejiri2,stack}.

\item
Polyakov loops written in terms of abelian fields and also 
in terms of Dirac strings of monopoles (monopole Polyakov loops) 
can reproduce the behavior of non-abelian Polyakov loops
\cite{suzu94a}.

\item
A monopole effective action can be calculated.
The argument of the energy and the entropy 
indicates that QCD is 
in the monopole condensed phase\cite{shiba3,shiba2a}.
\end{enumerate}

These facts are usually called abelian (monopole) 
dominance in quark confinement 
and suggest that 't Hooft's conjecture\cite{thooft} 
is realized in MA gauge.

Figure \ref{POLYAKOVLOOP} shows the non-abelian, the abelian and 
the monopole Polyakov loops versus $\beta$ on $24^3\times 4$ 
$SU(2)$ lattice\cite{suzu94a}.
The abelian and the monopole Polyakov loops change drastically 
around the critical point 
$\beta$=2.29 determined from the non-abelian Polyakov loops.
The abelian and the monopole Polyakov loops appear 
to be good order parameters.
However, those curves seem to have different slopes.
Their absolute values in the deconfinement phase are also different.
Actually, those three, the non-abelian, the abelian and the monopole 
Polyakov loops are quite different operators.

The critical property of 4-dimensional $SU(2)$ lattice gauge theory 
is shown to be universal to that of 3-dimensional $Z_2$ theory
\cite{ferr2}. It is interesting to study 
what exponents are calculated from the abelian 
and the monopole Polyakov loops at each critical point, 
since $Z_2$ symmetry is not so directly understood 
in the framwork of monopole dynamics\cite{ejiri} and 
 there is no reason the exponents of those different  
Polyakov loops agree with each other.
Such a study helps us also to test how good the abelian dominance is 
quantitatively. It is the aim of this work.

\section{Definition of abelian and monopole Polyakov loops } %----------
A non-abelian Polyakov loop in $SU(2)$ lattice gauge theory 
is written in the form
\begin{eqnarray}
 P_{SU(2)}(x_0)=\frac{1}{2}\mbox{Tr}\prod_t U_4(x_0,t)  ,
\end{eqnarray}
where $U_\mu (x,t)$ are $SU(2)$ link variables at 
space $x$ and at time $t$.

After abelian projection is over, we can define 
abelian Polyakov loops\cite{hio} 
written in terms of abelian link variables.
The abelian link variables can be separated from 
gauge-fixed link variables 
\begin{eqnarray}
\widetilde{U}_\mu (x,t)=C_\mu(x,t)u_\mu(x,t) ,
\end{eqnarray}
where $\widetilde{U}_\mu (x,t)$ is a gauge-fixed link variable. 
 $u_\mu(x,t)$ is a  diagonal matrix composed of phase factors 
of the diagonal components of $\widetilde{U}_\mu (x,t)$.
We can define an abelian Polyakov loop 
\begin{eqnarray}
 P_{abel}(x_0)=\mbox{Re\,}[\exp (i\sum_{x,t}\theta_4(x,t)J_4(x,t))] 
 \label{eqn:ABELPLOOP} .
\end{eqnarray}
Here $J_\mu(x,t)=\delta_{\mu,4}\delta_{x,x_0}$ and $\theta_\mu (x,t)$ are the 
angle variables of $u_\mu(x,t)$:
\begin{eqnarray}
u_\mu(x,t)=\matrxd{\exp(i\theta_\mu (x,t))}{0}
                  {0}{\exp(-i\theta_\mu (x,t))}.
\end{eqnarray}

The abelian Polyakov loop can be decomposed into two parts: 
a monopole part and a photon part\cite{suzu94a}.
An abelian field strength can be written as 
\begin{eqnarray}
  f_{\mu\nu}(x,t) = \partial_{\mu}\theta_{\nu}(x,t) -
    \partial_{\nu}\theta_{\mu}(x,t) ,
\end{eqnarray}
where $\partial_{\nu}$ is a forward derivative.
Rewriting this equation, we get
\begin{eqnarray}
    \theta_\mu (x,t) &=& - \sum_{x',t'} D(x-x',t-t')
    \partial'_{\nu}f_{\nu \mu}(x',t')
    -\sum_{x',t'} D(x-x',t-t')
     \partial_\mu \partial'_{\nu}\theta_{\nu}(x',t') %
    \label{eqn:phase} ,
\end{eqnarray}
where $\partial'_{\nu}$ is a backward derivative and 
$D(x,t)$ is a lattice Coulomb propagator which satisfies 
$\partial'_\mu\partial_\mu D(x,t)=-\delta_{x,0}\delta_{t,0}$.
Then the abelian Polyakov loop (Eq.(\ref{eqn:ABELPLOOP})) can be written 
in terms of the abelian field strength:
\begin{eqnarray}
  P_{abel}(x_0)=\mbox{Re\,}[\exp (-i\sum_{x,t,x',t'} D(x-x',t-t')
    \partial'_{\nu}f_{\nu 4}(x',t')J_4(x,t))]
    \label{eqn:ABELPLOOP2} .
\end{eqnarray}
Here the second term of Eq.(\ref{eqn:phase}) vanishes 
owing to  $\partial_4 J_4(x,t)=0$.
The abelian field strength can be separated into two parts:
\begin{eqnarray}
f_{\nu\mu}=\bar{f}_{\nu\mu}+2\pi n_{\nu\mu},
\end{eqnarray}
where $n_{\nu\mu}$ is an integer and $\bar{f}_{\nu\mu}\in [-\pi,\pi)$.
Then, rewriting Eq.(\ref{eqn:ABELPLOOP2}), we get
\begin{eqnarray}
  P_{abel}(x_0)&=&\mbox{Re\,}[
    \exp (-i\sum_{x,t,x',t'} D(x-x',t-t')
    \partial'_{\nu}\bar{f}_{\nu 4}(x',t')J_4(x,t))  \\
    & & \;\times
    \exp (-2\pi i\sum_{x,t,x',t'} D(x-x',t-t')
    \partial'_{\nu}n_{\nu 4}(x',t')J_4(x,t))]  \\
    &=& \mbox{Re\,}[P_1(x_0)\cdot P_2(x_0)] .
\end{eqnarray}
The monopole Polyakov loop, $P_{mono}(x_0)=\mbox{Re\,}P_1(x_0)$ is 
composed of Dirac strings of monopoles.
$P_{photon}(x_0)=\mbox{Re\,}P_2(x_0)$ only contains 
the contributions from photons.
Suzuki et al.\cite{suzu94a} have indicated that 
\begin{enumerate}
\item
$ P_{abel}(x_0)\sim P_{mono}(x_0)\times P_{photon}(x_0)$ in MA gauge.
\item
$P_{abel}(x_0)$, $ P_{mono}(x_0)$ and $P_{SU(2)}(x_0)$ vanish for 
$\beta < \beta_c$. 
\item
$P_{photon}$ is finite 
from $\beta=$2.1 to 2.5 and 
does not change drastically around the critical point. 
\end{enumerate}

\section {Finite-size scaling theory} %---------------------------------

We calculated the critical exponent of the non-abelian, 
the abelian and the monopole Polyakov loops 
from a finite-size scaling theory.
The singular part of the free energy density on 
$N_s^3\times N_t$ lattice has the following form:
\begin{eqnarray}
 f_s(x,h,N_s)=N_s^{-d}Q_s(xN_s^{1/\nu},hN_s^{(\beta+\gamma)/\nu}
,g_i N_s^{y_i}),
\end{eqnarray}
where $x=(T-T_c)/T_c$. Here the action contains the term $hN_s^{d}L$ 
($L$ denotes the magnetization) 
and $g_i$ are the irrelevant fields with exponent $y_i$.
By differentiating $f_s$ with respect to $h$ at $h=0$, we get 
\begin{eqnarray}
  \langle L\rangle(x,N_s)&=&N_s^{-\beta/\nu}Q_L (xN_s^{1/\nu},g_i 
N_s^{y_i}) ,\\
  \chi(x,N_s)&=&N_s^{\gamma/\nu}Q_\chi (xN_s^{1/\nu},g_i N_s^{y_i}) ,\\
  g_r(x,N_s)&=&Q_{g_r} (xN_s^{1/\nu},g_i N_s^{y_i}) ,
\end{eqnarray}  
where $\langle L \rangle$, $\chi$ and $g_r$ are 
order parameter, susceptibility and 4-th cumulant, respectively:
\begin{eqnarray}
  L&=&\frac{1}{N_s^3}\sum_xP(x)  ,\\
  \chi&=&N_s^3(\langle L^2\rangle - \langle L\rangle^2) , \\
  g_r&=&\frac{\langle L^4\rangle}{{\langle L^2\rangle}^2}-3.
\end{eqnarray}%
Expanding these equations with respect to $x$, we have at $x=0$
%% \begin{eqnarray*}
%%  L(x,N_s)&=&N_s^{-\beta/\nu}
%%  (c_0+(c_1+c_2)N_s^{-\omega}xN_s^{1/\nu}+c_3N_s^{-\omega}) \\
%%  \chi(x,N_s)&=&N_s^{\gamma/\nu}
%%  (c_0+(c_1+c_2)N_s^{-\omega}xN_s^{1/\nu}+c_3N_s^{-\omega}) \\
%%  g_r(x,N_s)&=&Q_{g_r} 
%%  (c_0+(c_1+c_2)N_s^{-\omega}xN_s^{1/\nu}+c_3N_s^{-\omega}) 
%%\end{eqnarray*}  
\begin{eqnarray}
  \langle L\rangle(x=0,N_s)&=&N_s^{-\beta/\nu}(c_0+c_3N_s^{-\omega}) %
  \label{eqn:Lomega} ,\\
  \chi(x=0,N_s)&=&N_s^{\gamma/\nu}(c_0+c_3N_s^{-\omega}) %
  \label{eqn:chiomega} ,\\
  g_r(x=0,N_s)&=& c_0+c_3N_s^{-\omega} , \label{eqn:grfit}
\end{eqnarray}
where we take only the largest irrelevant exponents ($-\omega$) 
into account.
We can calculate the critical point from the fit to the 
$N_s$-behavior of those equations.
The critical point can be defined as the point where a fit to the leading 
$N_s$-behavior has the least  $\chi^2$\cite{engels}.
Actually, the leading $N_s$-behavior of Eq.(\ref{eqn:Lomega}) 
and of Eq.(\ref{eqn:chiomega}) is given by
\begin{eqnarray}
  \ln \langle L\rangle(x=0,N_s)&=&-\frac{\beta}{\nu}\ln N_s + \ln c_0 
\label{eqn:Lfit} ,\\
  \ln \chi(x=0,N_s)&=&\frac{\gamma}{\nu}\ln N_s + \ln c_0 \label{eqn:chifit}.
\end{eqnarray}

From the fits to 
Eqs.(\ref{eqn:grfit}), (\ref{eqn:Lfit}) and (\ref{eqn:chifit}), 
we can find the position of the critical point $\beta_c$,
and obtain the values of $\beta/\nu$, $\gamma/\nu$ and $g_r^{\infty}$ 
at $\beta_c$ simultaneously.
Here $g_r^{\infty}$ is the value of $g_r$ on the infinite volume 
and is denoted by $c_0$ in Eq.(\ref{eqn:grfit}).
We also considered the derivatives of the observables 
with respect to $x$. 
The $N_s$-behavior of each derivative is given by 
\begin{eqnarray}
  \frac{\partial \langle L\rangle}{\partial x}(x,N_s) &=&
  N_s^{(1-\beta)/\nu}Q'_L(xN_s^{1/\nu}) \label{eqn:dLomega} ,\\
  \frac{\partial \chi}{\partial x}(x,N_s) &=&
  N_s^{(1+\gamma)/\nu}Q'_\chi(xN_s^{1/\nu}) \label{eqn:dchiomega} ,\\
  \frac{\partial g_r}{\partial x}(x,N_s) &=&
  N_s^{1/\nu}Q'_{g_r}(xN_s^{1/\nu}) \label{eqn:dgromega}.
\end{eqnarray}
The leading $N_s$-behavior of each equation at the critical point is 
\begin{eqnarray}
  \ln\frac{\partial \langle L\rangle}{\partial x}(x=0,N_s)
  &=& \frac{1-\beta}{\nu}\ln N_s + \ln d_0 \label{eqn:dLfit} ,\\
  \ln\frac{\partial \chi}{\partial x}(x=0,N_s)
  &=& \frac{1+\gamma}{\nu}\ln N_s + \ln d_0 \label{eqn:dchifit} ,\\
  \ln\frac{\partial g_r}{\partial x}(x=0,N_s)
  &=& \frac{1}{\nu}\ln N_s + \ln d_0 \label{eqn:dgrfit} .
\end{eqnarray}
Hence, 
$(1-\beta)/\nu$,$(1+\gamma)/\nu$ and $1/\nu$ 
can also be evaluated from the fits to those equations.

\section{Results and Discussions} %------------------------------------

We performed numerical calculations on 
$N_s^3\times 4$ lattices, where $N_s=$8, 12, 16 and 24.
The standard $SU(2)$ Wilson action was adopted and abelian link valuables 
were defined in MA gauge.
We calculated the observables
\begin{eqnarray}
  L&=&\frac{1}{N_s^3}\sum_xP(x)  ,
\end{eqnarray}%
where 
$P(x)$ denotes $P_{SU(2)}(x)$, $P_{abel}(x)$ and $P_{mono}(x)$. 
Actually, $\chi_v=N_s^3\langle L^2\rangle$ was calculated instead of $\chi$, 
because $\chi_v$ is equal to $\chi$ below the critical point \cite{engels}.
The values of the observables at various $\beta$ are needed 
in order to calculate the derivatives 
with respect to $x$, where $x=(\beta-\beta_c)/\beta_c$.
We used the density of state method(DSM)\cite{ferr,engels2}.
First we performed Monte-Carlo simulations at $\beta=\beta_0$,
and then calculated the following averages:
\begin{eqnarray}
O(s)=\sum_{k=1}^{N(s)}{O_k}/{N(s)} ,
\end{eqnarray}
where $s$ is the value of the action,
$N(s)$ the number of the configurations 
whose action has the same value of $s$, 
and $O_k$ the observable obtained from the $k$-th configuration.
The expectation value of the observables in the vicinity of $\beta_0$ 
is then given by 
\begin{eqnarray}
\langle O\rangle_\beta = \frac{\sum_s N(s)O(s)\exp(-\beta s+\beta_0 s)}
                 {\sum_s N(s)    \exp(-\beta s+\beta_0 s)}  ,
\end{eqnarray}
where $\beta_0=$2.2988 was adopted and $\beta \in$$[2.2980,2.3000]$. 
$\langle L\rangle$, $\chi$ and $g_r$ at $\beta_0$ were 
calculated every 50 sweeps 
after 2000 thermalization sweeps.
The number of samples was 100000,
except on $24^3\times 4$ lattice (47000 in the case).
The errors were determined according to the 
Jackknife method dividing the entire sample into 10 blocks 
(4 blocks on $24^3\times 4$ lattice).

\begin{figure}
\begin{center}
\leavevmode
  \parbox{80mm}{
  \epsfxsize=80mm\epsfysize=115mm\epsfbox{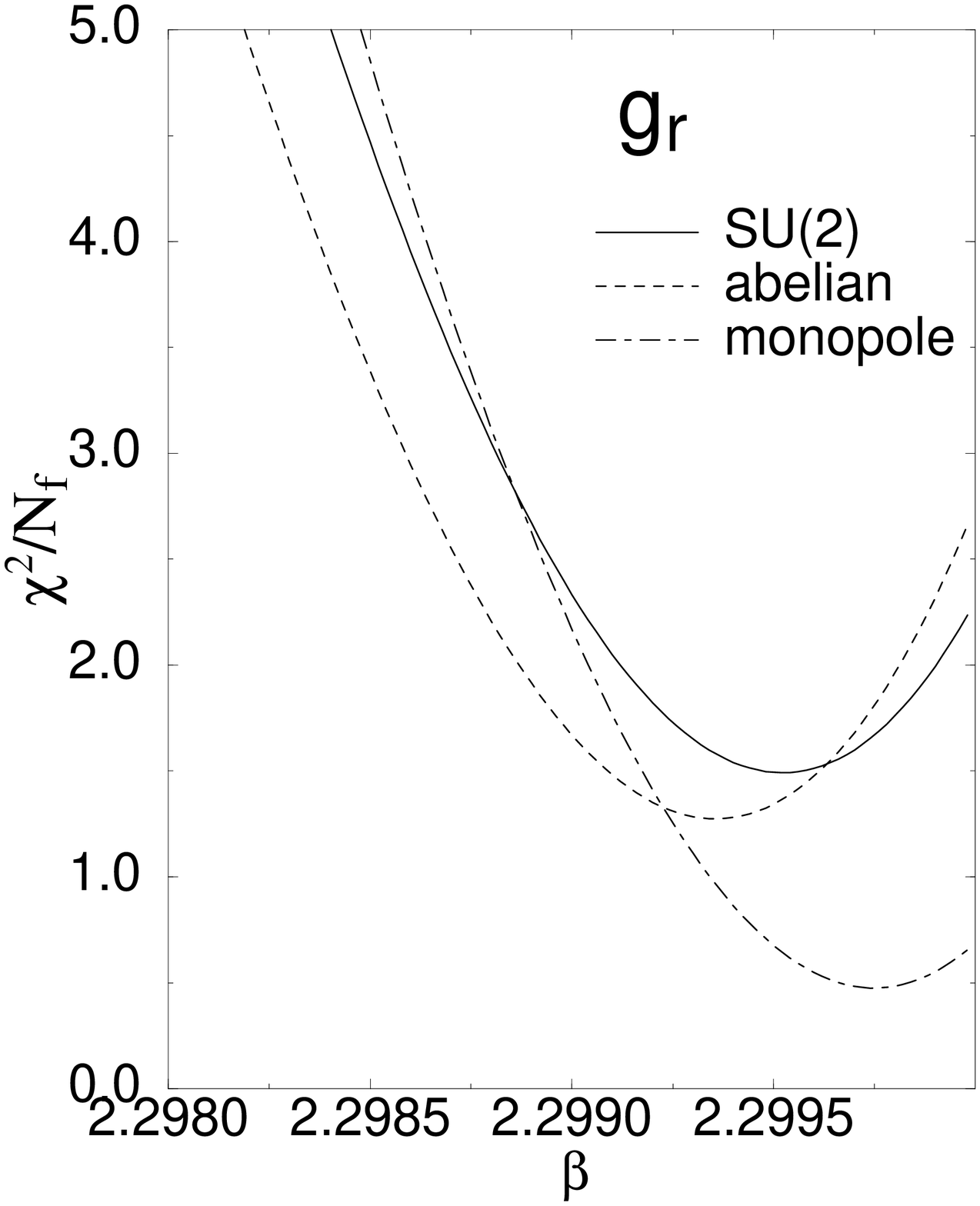}
  }
  \parbox{15mm}{}
  \parbox{80mm}{
  \epsfxsize=80mm\epsfysize=115mm\epsfbox{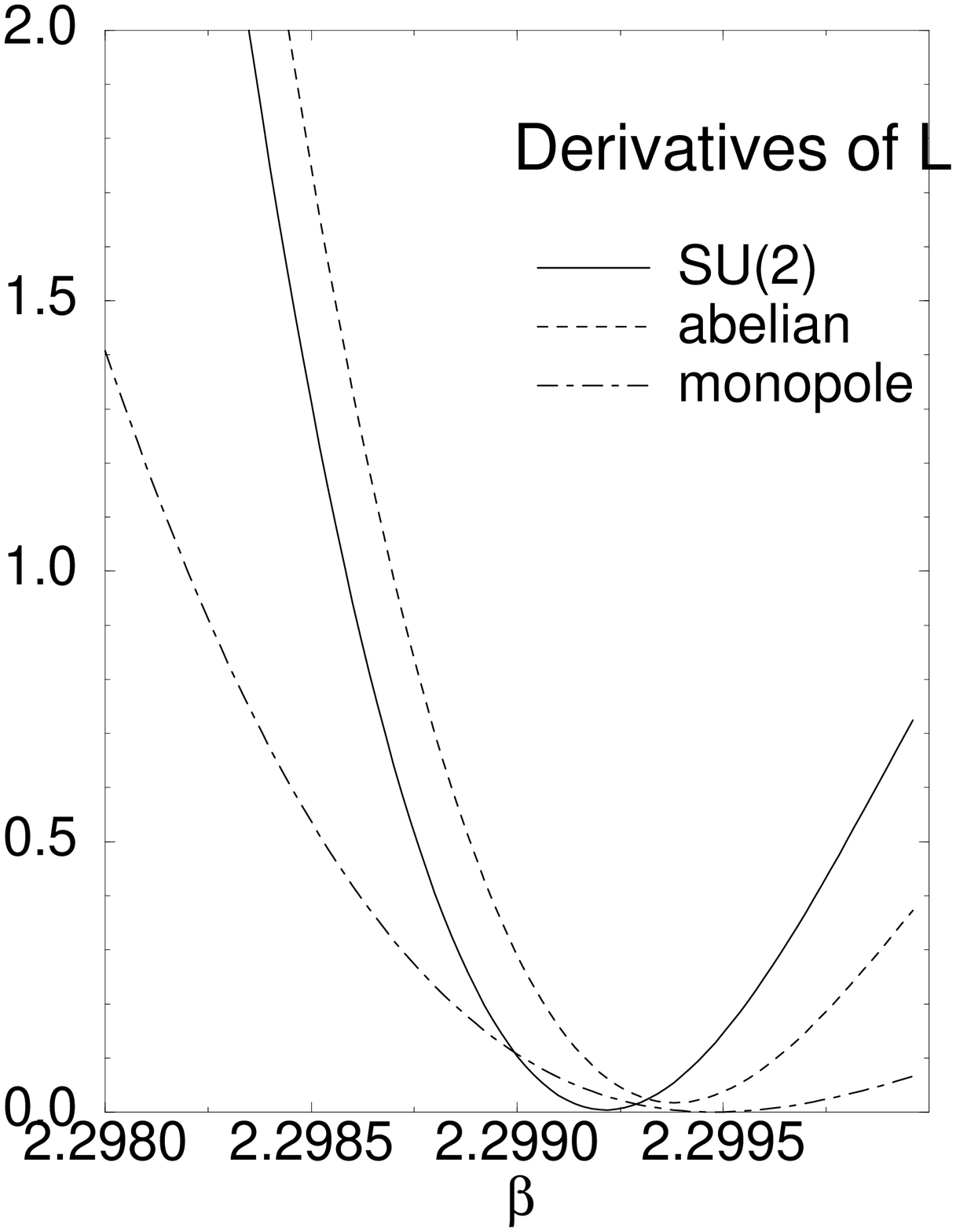}
  }
\end{center}
\caption{
$\chi^2/N_f$ from $g_r$ fits and from 
$\partial \langle L\rangle/\partial x$ fits 
versus $\beta$ in the non-abelian, the abelian and the monopole cases.
The number of degrees of freedom, $N_f$ is 2.
\label{CHISQ}
}
\end{figure}

We estimated the critical point $\beta_c$ from the $\chi^2$ method.
The data of our DSM results were fitted to 
Eqs.(\ref{eqn:grfit})-(\ref{eqn:chifit}) and 
Eqs.(\ref{eqn:dLfit})-(\ref{eqn:dgrfit}) 
at each $\beta$.
The number of input data was 2 and that of fit parameters was 2 
($\omega$ in Eq.(\ref{eqn:grfit}) was fixed to 1 in accordance 
with Engels et al.\cite{engels}).
Figure \ref{CHISQ} describes the typical curves of 
$\chi^2/N_f$ versus $\beta$.
Here the number of degrees of freedom, $N_f$ is 2.
Each curve in Fig. \ref{CHISQ} is smooth and has its minimum value.
Table \ref{table:CHISQMIN} shows the positions of minimal $\chi^2/N_f$ 
for all observables obtained.
Almost all the $\chi^2_{min}/N_f$ are small.
However, the $\chi^2_{min}/N_f$ were not seen from our 
$\partial g_r /\partial x$ fits.
Furthermore, our two-parameter fits were not so good 
in the cases of $\langle L\rangle$, $\chi$, 
$\partial \langle L\rangle/\partial x$ 
and $\partial \chi /\partial x$ 
from the monopole Polyakov loops. 
Then we used Eq.(\ref{eqn:Lomega}) and Eq.(\ref{eqn:chiomega}) 
for their fits 
which contained three parameters to be fitted.
The values of $\omega$ in Eq.(\ref{eqn:Lomega}) and Eq.(\ref{eqn:chiomega}) 
were chosen in such a way that the values of $\chi^2_{min}/N_f$ became 
as small as possible.

\begin{figure}
\vspace*{1cm}
 \epsfxsize=140mm
 \begin{center}
 \leavevmode
 \epsfbox{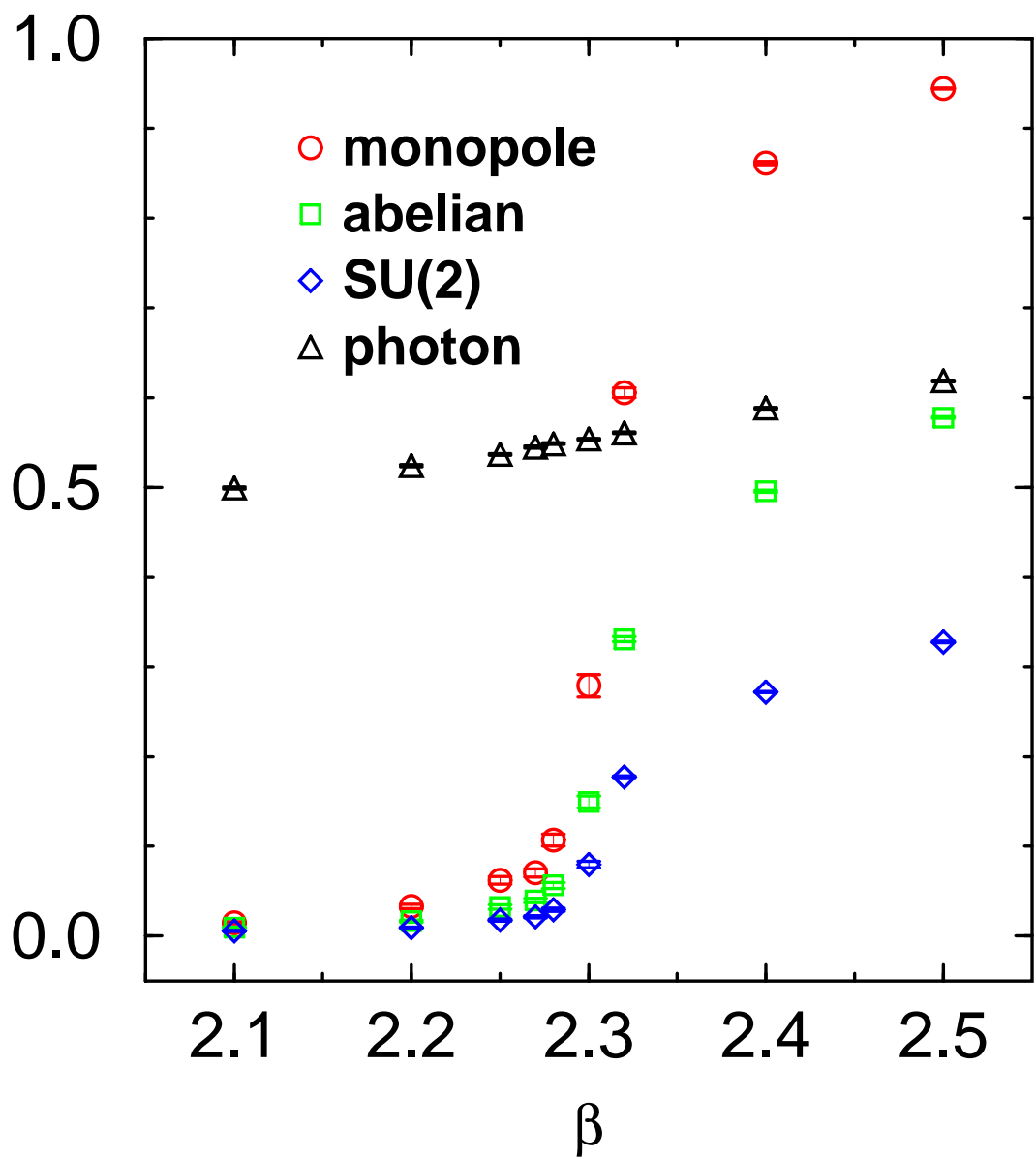}
 \end{center}
\caption{
Non-abelian, abelian and monopole Polyakov loops in $SU(2)$ QCD 
on $24^3\times 4$ lattices.
\label{POLYAKOVLOOP}
}
\end{figure}

Averaging 
the obtained minimal positions of $\chi^2_{min}/N_f$, we get
\begin{eqnarray}
   \beta_c^{SU(2)}      &=& 2.29940(20) , \\
   \beta_c^{abel} \;\;\,&=& 2.99962(26) , \\
   \beta_c^{mono} \;    &=& 2.29971(23) . 
\end{eqnarray}
The critical points obtained from the abelian and the monopole 
Polyakov loop are very close to the non-abelian critical point 
as expected from Fig. \ref{POLYAKOVLOOP}.

\begin{figure}
 \epsfxsize=140mm
 \begin{center}
 \leavevmode
 \epsfbox{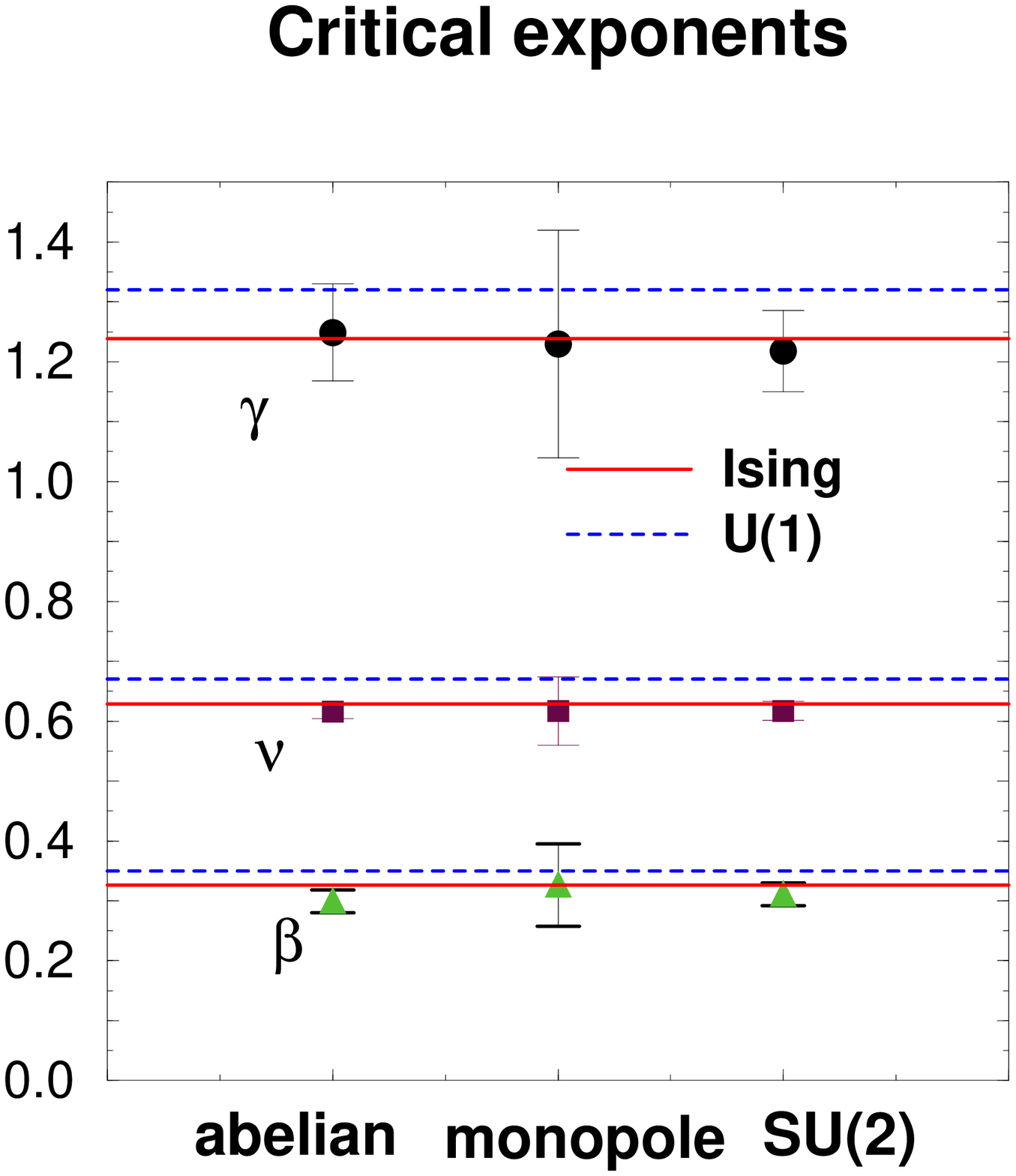}
 \end{center}
\caption{
Critical exponents of non-abelian, abelian and monopole Polyakov loops 
in $SU(2)$ QCD. 
\label{crexp}
}
\end{figure}

Table \ref{table:INDEX} lists the critical exponents on each critical point 
in the non-abelian, the abelian and the monopole case.
The non-abelian exponents obtained by Engels et al.\cite{engels}, 
the exponents of 3-dimensional Ising model\cite{ferr2} and 
those of $U(1)$\cite{svet} are also shown.
The errors were caused by fluctuations of the interpolated DSM data 
and by uncertainty of each $\beta_c$. See also Fig. \ref{crexp}.
Those critical exponents seem to be reliable 
because of the following reasons:
three $\nu$'s obtained from three different fits 
are within the statistical errors;
hyperscaling relations are well satisfied;
non-abelian exponents obtained are consistent with those 
of Engels et al.\cite{engels}.
Table \ref{table:INDEX} shows the following notable results:
\begin{enumerate}
\item
The critical exponents in the abelian and the monopole case 
are in agreement with non-abelian exponents within the statistical 
error.
\item
Those critical exponents agree with those of $Z_2$ rather than 
those of $U(1)$. 
\end{enumerate}
The abelian (monopole) dominance  
in quark confinement is proved quantitatively in this case.

There remain some problems to be studied further.
The critical points obtained are outside the one-sigma error bar 
of Engels et al.\cite{engels};
minimal $\chi^2/N_f$ were not seen for some fits;
some values of $\chi^2_{min}/N_f$ are $O(1)$ which are not small enough.
These problems seem to reflect a lack of statistics.
More samples may be needed especially on $24^3\times 4$ lattice.

The simulations of this work were carried out on VPP500 at 
Institute of Physical and Chemical Research (RIKEN) and 
at National Laboratory for High Energy Physics at Tsukuba (KEK).
This work is financially supported by JSPS 
Grant-in Aid for Scientific  
Research (B) (No.06452028).

%%%%%%%%%%%%%%%%% references %%%%%%%%%%%%%%%%%%%%%%%%%

%%%%%%%%%%%%% Table 1 %%%%%%%%%%%%%%%%%%
\begin{table}
\begin{center}
\begin{tabular}{|c|r|c|r||c|r|c|r|}
%  \hline
              &  & $\beta|_{\chi^2_{min}}$ & $\chi^2_{min}/N_f$ & %
              &  & $\beta|_{\chi^2_{min}}$ & $\chi^2_{min}/N_f$ \\
  \hline
  $g_r$       &  SU(2) & 2.29952 & 1.49 & %
  ${\partial g_r}/{\partial x}$&SU(2)&$-$ & $<O(10^{-1})$ \\
              &  abel & 2.29936 & 1.27 & %
                  &  abel & $-$  &  $<O(10^{-1})$ \\
              &  mono & 2.29974 & 0.47 & %
                  &  mono & $-$  &  $<O(10^{-1})$ \\
  \hline
  $\langle L\rangle$         &  SU(2) & 2.29960 & 1.33 & %
  ${\partial \langle L\rangle}/{\partial x}$& SU(2) & 2.29920 & 0.004 \\
              &  abel & 2.29984 & 0.95 & %
                  &  abel & 2.29938 & 0.017 \\
              &  mono & $-$     & $<O(1)$  & %
                  &  mono & 2.29948 & $6\times 10^{-7}$ \\
  \hline
  $\chi$      &  SU(2) & 2.29946 & 1.33 & %
  ${\partial \chi}/{\partial x}$& SU(2)  & 2.29924 & 0.005 \\
              &  abel & 2.29986 & 0.75 & %
                  &  abel & 2.29964 & 0.061 \\
              &  mono & $-$     &  $<O(1)$ & %
                  &  mono & 2.29992 & $6\times 10^{-6}$ \\
%  \hline
\end{tabular}
\end{center}
\caption{
The positions of minimal $\chi^2/N_f$ and the value of $\chi^2_{min}/N_f$ 
in the non-abelian,the abelian and the monopole cases.
\label{table:CHISQMIN}
}
\end{table}
%%%%%%%%%%%%% Table 2 %%%%%%%%%%%%%%%%%%
\begin{table}
\begin{center}
\begin{tabular}{|l|l|l|l||l|l||l|}
%\hline
 &{$SU(2)$}&{abel}&{mono}& {Engels et al.\cite{engels}} &
 {Ising\cite{ferr2}} & $U(1)$\cite{svet} \\  
\hline
$\beta/\nu$    & 0.504(18)& 0.485(22)& 0.528(64)& 0.525(8) & 0.518(7)& \\
$(1-\beta)/\nu$& 1.117(27)& 1.138(10)& 1.091(84)& 1.085(14)& 1.072(7)& \\
%%%- ${1}/{\nu}$   & 1.621(44)& 1.623(32)& 1.62(15) & 1.610(16)& 1.590(2) %
%%%- & \\
$\nu  $   & 0.617(16)& 0.616(12)& 0.617(57)& 0.621(6) & 0.6289(8)& 0.67 \\
$\beta$   & 0.311(19)& 0.299(19)& 0.326(69)& 0.326(8) & 0.3258(44)& 0.35 \\
\hline
$\gamma/\nu$   & 1.977(29)& 2.025(34)& 1.991(88)& 1.944(13)& 1.970(11) & \\
$(1+\gamma)/\nu$&3.600(38)& 3.646(44)& 3.608(93)& 3.555(15)& 3.560(11) & \\
%%%- ${1}/{\nu}$ & 1.623(67)& 1.621(78)& 1.62(18) & 1.611(20)& 1.590(2) %
%%%- & \\
$\nu  $   & 0.616(25)& 0.617(29)& 0.618(68)& 0.621(8) & 0.6289(8) & \\
$\gamma$  & 1.218(68)& 1.249(81)& 1.23(19) & 1.207(24)& 1.239(7)  & 1.32 \\
\hline
$\gamma/\nu+2\beta/\nu$ %
                  & 2.985(47)& 2.995(56)& 3.05(15)& 2.994(21)& 3.006(18)& \\
\hline
$-g_r^\infty$     & 1.447(41)& 1.438(42)& 1.438(41)& 1.403(16)& 1.41 & \\
%%%- ${1}/{\nu}$  & 1.579(33)& 1.611(38)& 1.668(38)& 1.587(27)& 1.590(2) %
%%%- & \\
$\nu$     & 0.633(13)& 0.621(14)& 0.600(13)& 0.630(11)& 0.6289(8) & \\
%\hline
\end{tabular}
\end{center}
\caption{
The critical exponents calculated from the non-abelian, the abelian and 
the monopole Polyakov loops at each critical point.
\label{table:INDEX}
}
\end{table}

%%%%%%%%%%%%%%%%%%%%% figures %%%%%%%%%%%%%%%%%%%%%%%%%%%

\end{document}